\documentclass[twocolumn]{revtex4}
\usepackage{amsmath}
\usepackage{amsfonts}
\usepackage{amssymb}
\usepackage{graphicx}
\usepackage[colorlinks,linkcolor=blue,anchorcolor=blue,citecolor=blue,urlcolor=black]%
{hyperref}
\usepackage{mathrsfs}
\usepackage{dcolumn}
\usepackage{bm}
\usepackage{epsfig}%
 \usepackage{graphicx}
\providecommand{\U}[1]{\protect\rule{.1in}{.1in}}

\def\be{\begin{equation}}
\def\ee{\end{equation}}
\begin{document}
\title{\bf Generating Spatial Spectrum with Metasurfaces}
\author{Wenwei Liu$^{1}$}
\author{Zhancheng Li$^{1}$}
\author{Hua Cheng$^{1}$}
\author{Chengchun Tang$^{3}$}
\author{Junjie Li$^{3}$}
\author{Shuang Zhang$^{2}$}
\email{s.zhang@bham.ac.uk}
\author{Shuqi Chen$^{1}$}
\email{schen@nankai.edu.cn}
\author{Jianguo Tian$^{1}$}
\email{jjtian@nankai.edu.cn}

\affiliation{$^{1}$The Key Laboratory of Weak Light Nonlinear Photonics, Ministry of Education, School of Physics and TEDA Institute of Applied Physics, Nankai University, Tianjin 300071, China.}
\affiliation{$^{2}$School of Physics and Astronomy, University of Birmingham, Birmingham B15 2TT, UK.}
\affiliation{$^{3}$Beijing National Laboratory for Condensed Matter Physics, Institute of Physics, Chinese Academy of Sciences, Beijing 100190, China}

\date{\today}

\begin{abstract}
Fourier optics, the principle of using Fourier Transformation to understand the functionalities of optical elements, lies at the heart of modern optics, and has been widely applied to optical information processing, imaging, holography \emph{etc}. While a simple thin lens is capable of resolving Fourier components of an arbitrary optical wavefront, its operation is limited to near normal light incidence, i.e. the paraxial approximation, which put a severe constraint on the resolvable Fourier domain. As a result, high−-order Fourier components are lost, resulting in extinction of high−-resolution information of an image. Here, we experimentally demonstrate a dielectric metasurface consisting of high−-aspect-−ratio silicon waveguide array, which is capable of performing Fourier transform for a large incident angle range and a broad operating bandwidth. Thus our device significantly expands the operational Fourier space, benefitting from the large numerical aperture (NA), and negligible angular dispersion at large incident angles. Our Fourier metasurface will not only facilitate efficient manipulation of spatial spectrum of free-space optical wavefront, but also be readily integrated into micro-optical platforms due to its compact size.
\end{abstract}

\maketitle

Fourier optics represents an important platform for diverse applications such as, spatial filtering \cite{1,2}, compressed sensing \cite{3}, and holography \cite{4,5,6}. The basic building block of Fourier optics is Fourier lens, which can perform a Fourier transform of the incident wavefront at the focal plane \cite{7}. Traditional Fourier lens is realized by a thin transparent medium with slowly varying thickness that can accumulate optical phase under the paraxial approximation. However, this approximation greatly restricted the ability of Fourier lens to resolve a large \emph{k} component of a wavefront. Thus far, there has been no demonstration of a material or a design that works as a Fourier lens beyond the paraxial approximation, both in conventional optics and nano-optics.

In recent years, the burgeoning field of metasurfaces provides a high degree of freedom for locally tailoring the optical properties on a subwavelength scale based on plasmonic \cite{8,9,10,11,12} and dielectric building blocks \cite{13,14,15,16}. Complete control over the phase and polarization of light has been achieved via dielectric metasurfaces \cite{17}. With delicate design of phase distribution, 3D holography \cite{18}, multi−wavelength achromatic holography \cite{19}, and nonlinear holography \cite{20} have also been accomplished. Metasurfaces can be designed to operate as flat lenses, the so called metalens. With a tailored hyperbolic phase distribution, metalens can convert a plane wavefront to a spherical one \cite{21,22,23,24}. This novel method has been applied to cancel chromatic aberration in the infrared waveband \cite{25,26} and the visible regime \cite{27} via judicious phase design. It is also demonstrated that metalens can replace conventional objectives for sub−wavelength resolution imaging \cite{28}. Recently, a doublet corrected metalens is proposed to achieve monochromatic aberrations with an incident angle reaching $30^{\circ}$ \cite{29}, which is approaching the boundary of the paraxial approximation ($\sin \pi /6 \approx \pi /6$). However, the abovementioned metalenses are designed for imaging and cannot be utilized to perform Fourier transform, especially when the incident angle is greater than $30^{\circ}$. Here, we demonstrate a Fourier metalens made of an array of dielectric waveguide resonators, which shows good performance for $0-60^{\circ}$ of incidence angle covering a broad bandwidth from 1,100 to 1,700 nm. The superior performance of the metalens is confirmed through the Fourier transform / Fraunhofer diffraction of a grating with periodicity slightly greater than the wavelength. The platform we propose does not suffer from the paraxial approximation and provides a powerful framework for realizing optical Fourier operation.

\begin{figure*}[tb]
\centerline{\includegraphics[width=17cm]{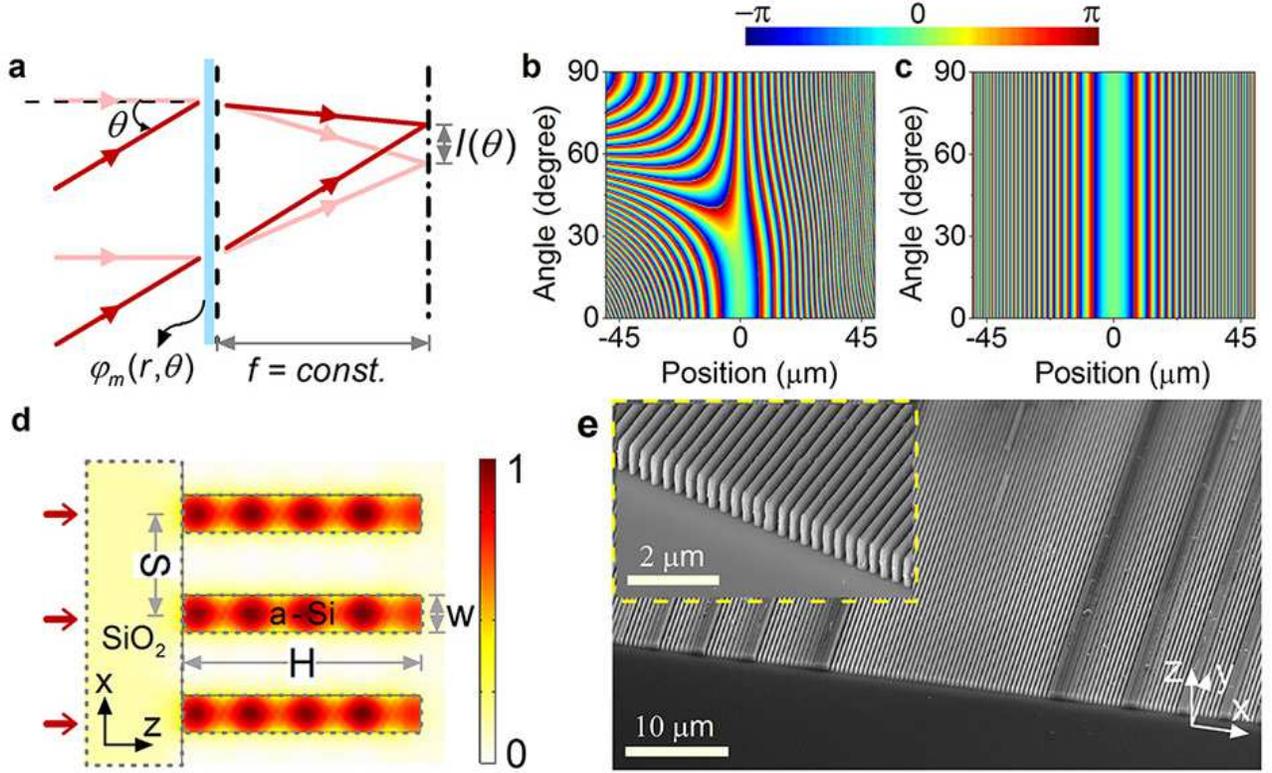}}
\caption{\textbf{Design and fabrication of the Fourier metalenses.} \textbf{a}, With an incident angle of $\theta$, the focal spot will move a distance of $l(\theta)$ along the focal plane. \textbf{b}, The required phase delay against positions of the nanostructures and the incident angle, where $l(\theta ) = f\sin \theta $ (Fourier lenses). \textbf{c}, The angle-dispersion-free phase delay with a first order approximation, against positions of the nanostructures and the incident angle. \textbf{d}, Schematic of the Fourier metalens and its building block. The amorphous silicon nanolines are designed with: H = 1,050, S = 450 nm. \textbf{e}, SEM micrograph of the fabricated Fourier metalens.}
\label{fig1}%
\end{figure*}

A generic illustration of the light path for oblique focusing of monochromatic light is shown in Fig. \ref{fig1}a. Each incident angle $\theta$ corresponds to a focus point in the focal plane, with an foci offset of $l(\theta)$ from the origin, which is called spatial frequency in Fourier optics. To achieve the desired focusing functionality, an abrupt phase shift $\varphi_m$ is required along the surface of the metasurface to compensate the phase accumulated via propagation:
\begin{align}
{\varphi _m}(r,\theta ) =  &- k[r\sin \theta  + \sqrt {{{(r - l(\theta ))}^2} + {f^2}} \nonumber \\
&- \sqrt {l{{(\theta )}^2} + {f^2}}]
\label{eq:e1}
\end{align}

When $\theta$ and $l(\theta)$ are both equal to zero, Eq. (\ref{eq:e1}) recovers the hyperbolic formula working in normal incidence \cite{21}. For a Fourier lens, $l(\theta ) = f\sin \theta $ is required, and the corresponding phase shift $\varphi_m$ is computed via Eq. (\ref{eq:e1}) and shown in Fig. \ref{fig1}b. With an incident angle above $30^{\circ}$, $\varphi_m$ is highly angle-dispersive, which hinders the development of large-angle Fourier devices. As a comparison, the foci offset for an imaging lens is $l(\theta ) = f\tan \theta $ (Supplementary Fig. 1). Thus, the existing high numerical aperture lenses with larger field-of-view \cite{28,29} are not applicable as Fourier lenses. Another type of lens for manipulating the focal plane is F-Theta Scan Lens with $l(\theta ) = f \theta $, which is often used in carving systems and scanning systems. Commercial F-Theta Scan Lenses are usually designed to work at small incident angles, such as Thorlabs FTH100 ($\pm 28^{\circ}$), because $\varphi_m$ is also highly angle-dispersive for large incident angles. Generally, an artificial design of $l(\theta)$ is required to match various needs of manipulating the focal plane. Here we demonstrate a Fourier metalens breaking the paraxial condition based on dielectric waveguides (DWs). By applying a Taylor series ${(1 + x)^a} \approx 1 + ax$ to the first order, Eq. (\ref{eq:e1}) can be simplified as:
\begin{align}
{\varphi _m}(r,\theta ) \approx  - kr(\sin \theta  + \frac{{r - 2l(\theta )}}{{2f}})
\label{eq:e2}
\end{align}
Substituting $l(\theta ) = f\sin \theta $ into Eq. (\ref{eq:e2}), we obtain ${\varphi _m}(r,\theta ) \approx  - k{r^2}/2f$ , which is angle-dispersion-free. Furthermore, since it is the phase difference that determines the wavefront of output light rather than the absolute phase value, the phase equation can be generalized as:
\begin{align}
{\varphi _m}(r,\theta ) \approx  - k{r^2}/2f + \varphi (0,\theta )
\label{eq:e3}
\end{align}
where $\varphi(0,\theta)$ is the phase at zero point under an oblique incident angle $\theta$. The term $\varphi(0,\theta)$ relaxes the constraint in Eq. (\ref{eq:e2}) and simplifies the realization of an angle-dispersion-free phase distribution (Supplementary Figs. 7−-9). The desired phase distribution according to Eq. (\ref{eq:e3}) is computed and shown in Fig. \ref{fig1}c. To achieve this phase distribution, we designed DWs of amorphous silicon (a-Si) on fused silica substrate (Fig. \ref{fig1}d). Figure \ref{fig1}e shows the scanning electron microscope (SEM) images of the metalens fabricated using the a-Si DWs. The size of the designed Fourier metalens is $190 \times 100$ $\mu m$ (190 $\mu m$ along $x$ direction, for details of fabrication and measurement, see Methods).

\begin{figure*}[t]
\centerline{\includegraphics[width=17cm]{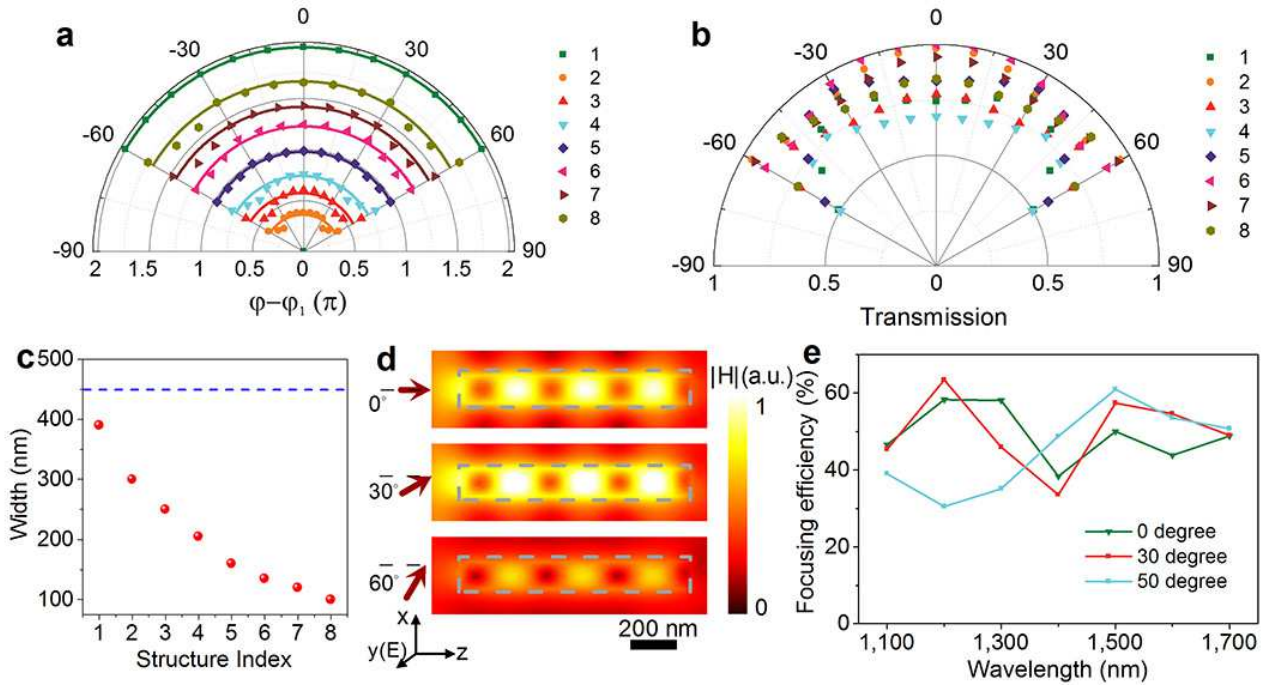}}
\caption{\textbf{Theoretically realizing the Fourier metalens.} \textbf{a}, Simulated phase difference $\varphi-\varphi_1$ (dots) for selected eight DWs under $-60^{\circ}$ to $60^{
\circ}$ incidence, where $\varphi$ is the phase of each DW, and $\varphi_1$ is the phase of the first DW. The solid lines are eye guides. \textbf{b}, Simulated transmission of each DW under $-60^{\circ}$ to $60^{\circ}$ incidence. All of the transmission are larger than 0.5, and have an average value at around 0.8. The incident wavelength in $\textbf{a}-\textbf{b}$ is 1,550 nm. \textbf{c}, Eight widths of the nanolines (red dots) in the simulation and experiments. All of these widths are smaller than the size of the unit cell (blue dashed line) by at least 60 nm to avoid adhesion of DWs. \textbf{d}, Simulated magnetic field intensity (1,550 nm) when light propagates at $0^{\circ}$, $30^{\circ}$, and $60^{\circ}$, respectively. The width of the DW is 160 nm with structure index of 5. \textbf{e}, Simulated focusing efficiency as a function of wavelength. This efficiency is defined as the fraction of the incident optical power that is converted to the transmitted focusing light.}
\label{fig2}%
\end{figure*}

We select eight widths of DWs using a commercial finite element method-based COMSOL Multiphysics software \cite{30} (Fig. \ref{fig2}c). Permittivity of a-Si used in the simulation is taken from Ref. \cite{31}. The simulated phase difference $\varphi-\varphi_1$ at 1,550 nm is shown in Fig. \ref{fig2}a, where $\varphi_1$ is the phase of the structure with the index of one (Fig. \ref{fig2}c). According to the simulation, the eight DWs have a phase variation in the whole range $0-2\pi$, and the phase variation for each DW is less than $0.164\pi$ under $0-60^{\circ}$ incidence. This phase profile satisfies Eq. (\ref{eq:e3}) for achieving Fourier operation. The simulated transmission $\left| t \right|$ is around 0.8, and the minimum $\left| t \right|$ with the fourth DW is over 0.5 (Fig. \ref{fig2}b). The simulated magnetic field intensities with incident angle $\theta$ of $0^{\circ}$, $30^{\circ}$, and $60^{\circ}$ in Fig. \ref{fig2}d (the fifth DW with width of 160 nm) show that light is mainly confined inside the high-refractive-index DWs, with negligible coupling between the neighboring DWs, and the scattered light is primarily determined by the transmission and phase of each DW \cite{32}. Furthermore, the waveguide mode remains nearly the same under different incident angles, i.e. the angle-dispersion-free performance is satisfied. In a simplified picture, each DW can be treated as a low-quality-factor Fabry-P\'erot resonator, with the effective refractive indices of the waveguide determined by the width of the waveguide. Thus, full control of phase can be achieved through optimizing the width of DWs, and transmission is determined by the quality-factor of the waveguide. Simulations in Fig. \ref{fig2}e show that focusing efficiencies as high as about $50\% $ are achieved through the waveband of 1,100--1,700 nm. The focusing efficiency is calculated as the ratio of focused power to the total incident optical power.

\begin{figure*}[t]
\centerline{\includegraphics[width=17cm]{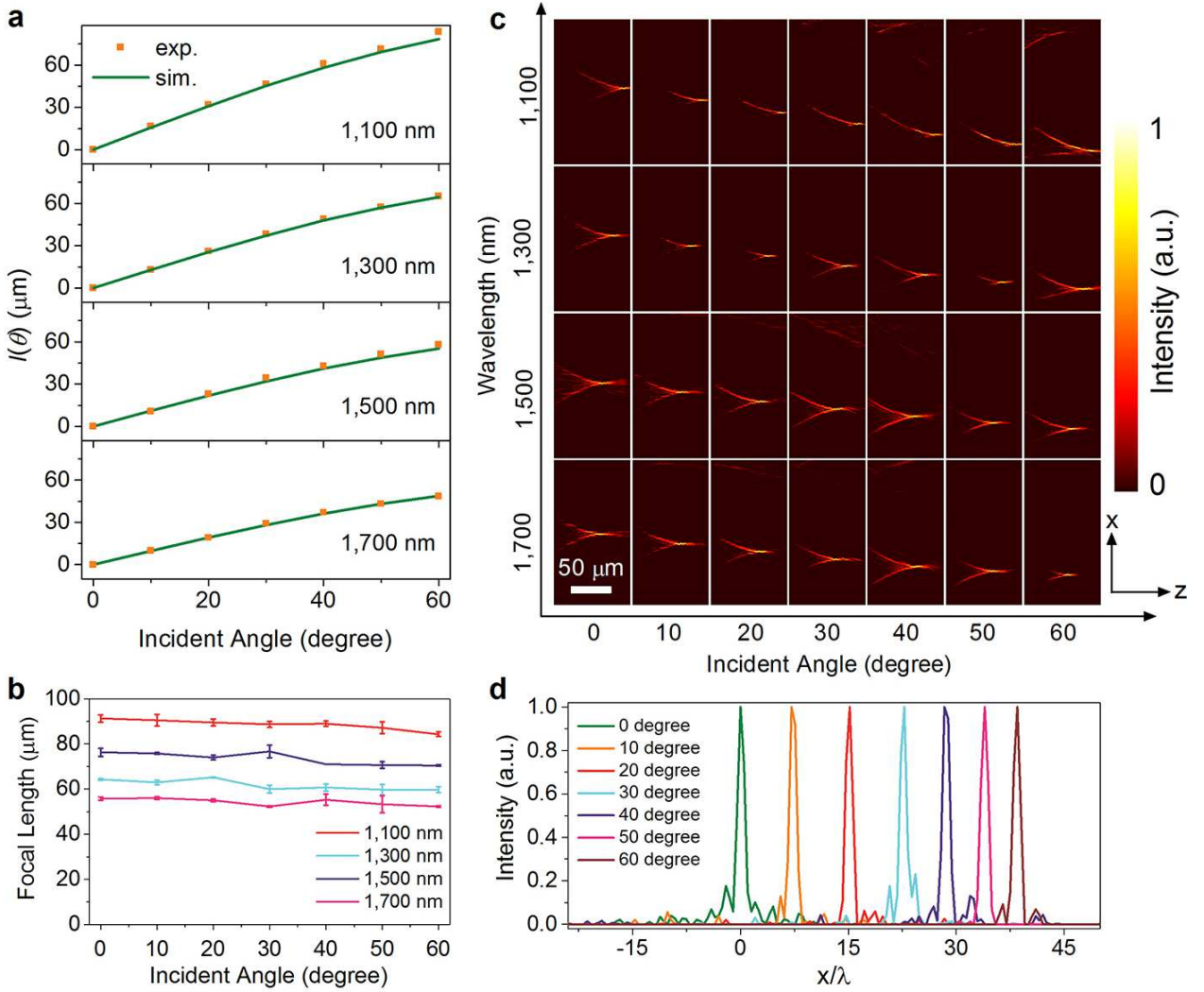}}
\caption{\textbf{Experimental demonstration of the Fourier metalens with incline incidence.} \textbf{a}, \textbf{b}, Measured and simulated (\textbf{a}) foci offset $l(\theta)$ and (\textbf{b}) focal length with different incident wavelengths and angles. \textbf{c}, Intensity distribution of the $x$-$z$ plane, showing the evolution of the foci offset $l(\theta)$ for different incident angles and wavelengths. \textbf{d}, Measured focusing vertical cut (1,500 nm) at 0 to $60^{\circ}$ incidence.}
\label{fig3}%
\end{figure*}

As shown in Fig. \ref{fig3}a, the measured foci offset $l(\theta)$ shows a good agreement with the theoretical one: $l(\theta ) = f\sin \theta $ across a broad bandwidth. The measured focal lengths as a function of incident angles for different incident wavelengths are shown in Fig. \ref{fig3}b. Since the metalens has a chromatic dispersion, the focal length decreases as the wavelength increases. Importantly, there is no field curvature in the focal plane since the focal length remains almost the same with varied incident angles. This characteristic fits the design schematic shown in Fig. \ref{fig1}a, consequently the focal plane is a flat rather than a curved surface, which facilitates the spatial filtering process in applications of Fourier optics \cite{7}. As shown in Supplementary Fig. 10, the simulated focal lengths remain the same for $0-60^{\circ}$ incidence. In addition, we measure the beam intensity profile of the metalens in the $x-z$ cross section (Fig. \ref{fig3}c), showing the evolution of the foci offset $l(\theta)$ for different incident wavelengths and angles. The weak noises in the profiles are caused by the bright field image of incident light, which, nonetheless, does not affect the measurement of $l(\theta)$ and focal length. The intensity profiles of the vertical cuts of the focal lines for 1,500 nm is shown in Fig. \ref{fig3}d. For $0-60^{\circ}$ incidence, full width at half maximum (FWHM) is about $\lambda$ and an evident shift is observed along $x$-axis when increasing the incident angle.

\begin{figure*}[t]
\centerline{\includegraphics[width=17cm]{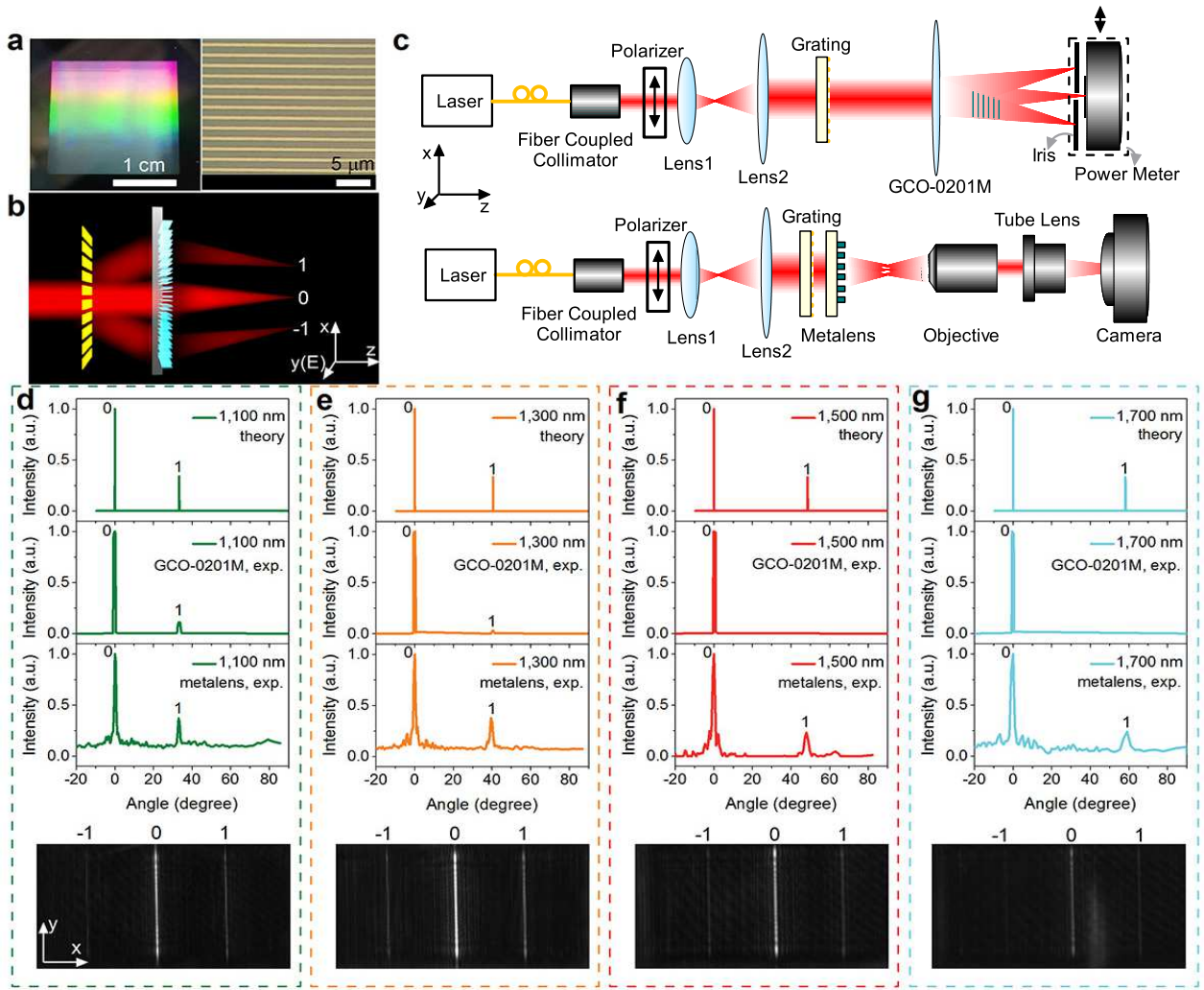}}
\caption{\textbf{Fourier transform of a grating by the metalens.} \textbf{a}, Photograph of the grating under a fluorescent lamp (left). Photograph of the grating taken by a CCD camera (right). The grating has a period of 2 $\mu m$, and duty cycle of 0.45. \textbf{b}, Schematic of the Fourier transform via the metalens. \textbf{c}, Illustration of the measurement set-up for the Fourier transform through a commercial Fourier lens GCO-0201M (top) and the metalens (bottom). \textbf{d} to \textbf{g}, Fourier transform in the focal plane at wavelengths of (\textbf{d}) 1,100 nm, (\textbf{e}) 1,300 nm, (\textbf{f}) 1,500 nm, (\textbf{g}) 1,700 nm. Top (\textbf{d}−-\textbf{g}): The $0^{\rm th}$ and $1^{\rm st}$ order diffraction according to the theoretical calculation and the corresponding measurement through GCO-0201M and the metalens. Bottom (\textbf{d}−-\textbf{g}): photographs of the Fourier-transformed focal plane through the metalens taken by an InGaAs camera.}
\label{fig4}%
\end{figure*}

\begin{figure*}[t]
\centerline{\includegraphics[width=17cm]{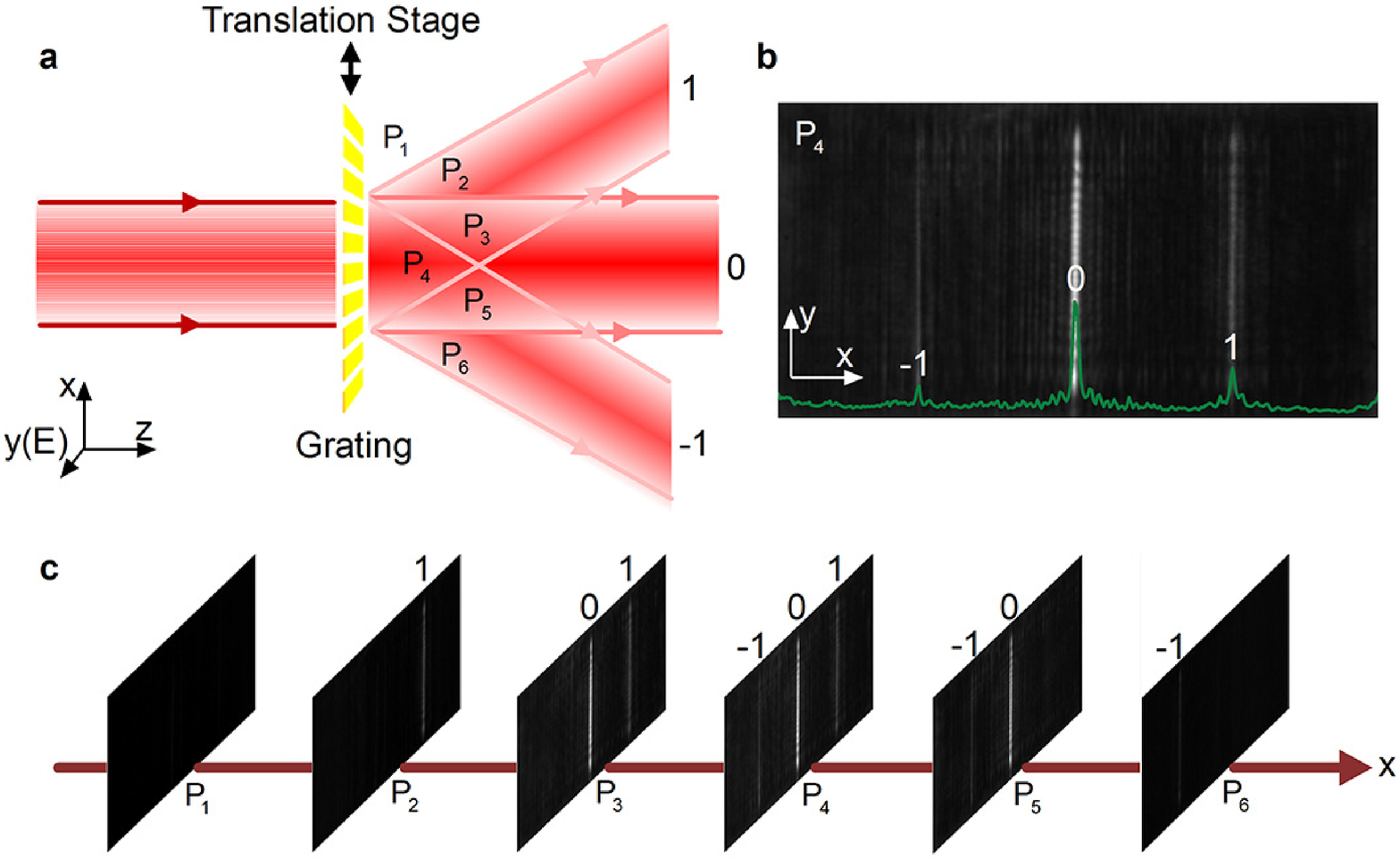}}
\caption{\textbf{Spatial filtering of the Fourier metalens in the focal plane.} \textbf{a}, schematic of the filtering process. By moving the grating along the $x$-axis through a translation stage, the metalens will locate at the positons of $P_1$--$P_6$, and the metalens will transform the incident light to various spatial spectra. \textbf{b}, Intensity distribution in the focal plane for 1,300 nm incidence with the metalens at $P_4$, where no spatial filtering occurs. \textbf{c}, Intensity distribution with the metalens located at the positions of $P_1$--$P_6$ (1,300 nm). The diffraction orders of ${-1}$, $0$, $1$ in sequence arise and vanish, mimicking spatial filtering through a rejector.}
\label{fig5}%
\end{figure*}

To demonstrate the Fourier transform ability of the metalens, we design a transmission grating as shown in Fig. \ref{fig4}a. The grating constant is 2 $\mu m$, allowing $0^{\rm th}$ and $\pm 1^{\rm st}$ order of Fraunhofer diffraction from 1,100--1,700 nm \cite{7}. The schematic of the Fourier transform process is shown in Fig. \ref{fig4}b. Since the grating is a dispersive device, the longer the incident wavelength is, the larger the diffraction angle will be. The diffraction angle is characterized by using the set-up shown in Fig. \ref{fig4}c. The grating is illuminated by $y$-polarized light, and the diffraction field is Fourier-transformed by a commercial Fourier lens and the metalens, respectively. As shown in the top row of Fig. \ref{fig4}d--g, the theoretical calculated divergent angle of the $1^{\rm st}$ order diffraction is $33.4^{\circ}$ for 1,100 nm, $40.5^{\circ}$ for 1,300 nm, $48.6^{\circ}$ for 1,500 nm, $58.2^{\circ}$ for 1,700 nm, respectively. When using the metalens to perform the Fourier transform, the measured angle of the $1^{\rm st}$ order diffraction is $33.0^{\circ}$ for 1,100 nm, $39.4^{\circ}$ for 1,300 nm, $48.0^{\circ}$ for 1,500 nm, and $59.0^{\circ}$ for 1,700 nm, which show a good agreement with the theoretical calculation. In comparison, as the commercial Fourier lens is designed under the paraxial approximation, it cannot work for large incident angles (larger than $30^{\circ}$). Although the commercial Fourier lens can exhibit the $1^{\rm st}$ order diffraction for the wavelengths of 1,100 nm and 1,300 nm, which is $33.0^{\circ}$ and $40.0^{\circ}$ respectively, the intensity of the $1^{\rm st}$ order diffraction is about an order of magnitude less than the theoretical ones. Whereas the measured intensities of the $1^{\rm st}$ order diffraction by the metalens are very close to the theoretical ones. The measured Fraunhofer diffraction spectra are obtained by $\theta  = \arcsin (z/f)$, where $z$ is the $z$-coordinate of the captured field profile in the focal plane, and $f$ is the measured focal length. The measured diffraction peaks are broader than the calculated ones, which can be attributed to the limited size of the grating and efficiency of the metalens. The captured intensity profiles by an InGaAs camera in the focal plane are shown in the bottom row of Fig. \ref{fig4}d-g, displaying $0^{\rm th}$ and $\pm 1^{\rm st}$ order diffraction for each incident wavelength. Furthermore, we demonstrate spatial filtering using the grating and the metalens (1,300 nm). By moving the grating along $x$-direction through a translation stage, the metalens is illuminated in different positions from $P_1$ to $P_6$, as shown in Fig. \ref{fig5}a. Taking $P_3$ as an example, the metalens is illuminated by the $0^{\rm th}$ and $1^{\rm st}$ order diffraction, with the $-1^{\rm st}$ order being filtered out. The intensity distribution without spatial filtering is shown in Fig. \ref{fig5}b, and the filtering process from $P_1$ to $P_6$ can be seen in Fig. \ref{fig5}c. The change in the recorded pattern with moving the grating is shown in Supplementary Movie 1 to 4 for different incident wavelengths (1,100, 1,300, 1,500, 1,700 nm). The spatial filtering results further confirm that the designed Fourier metalens is capable of performing Fourier transform of the incident wavefront at large incident angles.

Our work demonstrates that, through the coupling of the incident light with DW resonators, a near angle-dispersion-free phase change can be achieved. By choosing the widths of DWs, a Fourier metalens allowing large incident angles is demonstrated and agrees well with the theoretical predictions. Considering the broadband behavior, our metalens is compatible with commonly used telecommunication platform, could be readily integrated into micro-optical platforms, and could directly replace conventional thin lenses. We note that the theoretical approach and the design technique can be simply introduced to similar platforms, such as titanium dioxide based metalens in the visible wavelengths \cite{33}. This approach can also be extended to 2D or 3D Fourier devices to exhibit a full reciprocal lattice in $k$-space, which introduces more degrees of freedom for describing arbitrary electromagnetic waves, such as Bessel beams and Airy beams.

\begin{center}
\textbf{Methods}
\end{center}

\textbf{Simulations and design.}

The simulated results in Fig. \ref{fig2} and \ref{fig3}a are obtained through an FEM simulation via COMSOL Multiphysics. The permittivity of amorphous silicon is taken from Ref. \cite{31}, and the substrate is fused-silica with refractive index of 1.45. During all the simulations, light is illuminated from the substrate side. According to the equation of grating ${k_{i\parallel }} + m{k_\Lambda } = {k_{o\parallel }}$ (${k_{i\parallel }}$/${k_{o\parallel }}$ is the incident/output wavevector projected to a direction parallel to the metalens, and ${k_\Lambda }$ is the reciprocal lattice vector of the metalens), the array is non-diffractive in air/substrate when the incident wavelength is larger than 839.7/1,217.6 nm for $60^{\circ}$ incidence. The calculated wavelength limits are also suitable for smaller incident angles. The minimum working wavelength is 1,100 nm in our design, which is really close to 1,217.6 nm, and the simulations and experiments both show a low influence by the high-order diffraction.

\textbf{Sample fabrication.}

The metalens were fabricated on a fused silica substrate. Firstly, a layer of 1,050 nm amorphous silicon is coated on the substrate with plasma enhanced chemical vapor deposition (PECVD) method. Then a layer of PMMA with thickness of 200 nm and a layer of Pedot:PSS with thickness of 35 nm were spun coated on the substrate by sequence. The Pedot:PSS layer was intended for charge release during the electron-beam lithography (EBL) process which employing a 100 kV voltage, 200 pA current and 1,000 $\mu C/cm^2$ dose. After the EBL process, the Pedot:PSS layer was removed with pure water for 60 s and PMMA was developed with MIKE/IPA(3:1) for 40 s. Then a 50 nm Cr layer was deposited on the resist with electron beam evaporation deposition (EBD) method. Then the Cr film was striped by removing PMMA with hot acetone at degree of $60^{\circ}C$ for 20 min, which is followed by HBr plasma dry etching with ICP machine. The residue Cr resist was remove by wet etching method. Finally, 3 nm Cr layer was coated for SEM and then removed with wet etching method before optical measurement.

\textbf{Measurement procedure.}

The measurements in Fig. \ref{fig3} are based on a home-built microscopy as shown in Supplementary Fig. 6a. The light source is a supercontinuous laser (NKT SuperK EXR-20), and the laser beam is collimated by a fiber collimator. Then the collimated beam passes through an infrared polarizer, and the output light from the sample is collected with an objective (Sigma near plan apo $50\times$, NA = 0.67), a tube lens (Thorlabs TL200-3P) and an InGaAs camera (HAMAMATSU InGaAs C10633). The objective, tube lens and InGaAs camera are all integrated on a XYZ translation stage to scan the focusing profile of the metalens.
The measurements in Fig.\ref{fig4} and \ref{fig5} are accomplished with the optical setup in Fig. \ref{fig4}c. We use a pair of lenses (focal length of 30 mm and 150 mm) to magnify the beam diameter up to about 5 mm, making the light diffracted by the grating illuminate the sample for all incident wavelengths. The grating is placed in front of the metalens with a distance of about 1 mm. With a normal incidence, the photography is captured in focal plane with the InGaAs camera. The spatial filtering measurement is via moving the grating along $x$-axis through a translation stage. We also use a commercial Fourier lens GCO-0201M (f = 300 mm) to compare with our metalens. A power meter (Thorlabs S132C) mounted with an iris is located at the focal plane of the Fourier lens. The diffraction angle is calculated through $\arcsin (z/f)$, where $z$ is the measured position of the power meter, and $f$ is the focal length.

\textbf{ACKNOWLEDGMENTS}

This work was supported by the National Key Research and Development Program of China (2016YFA0301102), the Natural Science Foundation of China (11574163 and 61378006), and the Program for New Century Excellent Talents in University (NCET-13-0294). We also acknowledge the support from the Collaborative Innovation Center of Extreme Optics, Shanxi University, Taiyuan, Shanxi 030006, China.

\textbf{Author contributions}

W.L. and S.C. initiated the idea. W.L., S.C., Z.L., and H.C. performed the numerical simulations and experiments. C.T. and J.L. fabricated the samples. W.L., S.C., S.Z., and J.T. prepared the manuscript. S.C. supervised the project. All the author discussed and analyzed the results.

\textbf{Competing financial interests}

The authors declare no competing financial interests.

\end{document}